\documentstyle[12pt,epsfig]{article}

\newcommand{\agt}{\,\rlap{\lower 3.5 pt \hbox{$\mathchar \sim$}} \raise 1pt
 \hbox {$>$}\,}
\newcommand{\alt}{\,\rlap{\lower 3.5 pt \hbox{$\mathchar \sim$}} \raise 1pt
 \hbox {$<$}\,}
\newcommand{\arsinh}{\mathop{{\mbox{arsinh}}}\nolimits}
\newcommand{\arcosh}{\mathop{{\mbox{arcosh}}}\nolimits}
\newcommand{\re}{\mathop{{\mbox{Re}}}\nolimits}

\begin{document}

\title{\vskip-3cm{\baselineskip14pt
\centerline{\normalsize\hfill MPI/PhT/98--054}
\centerline{\normalsize\hfill hep--ph/9807480}
\centerline{\normalsize\hfill July 1998}
}
\vskip1.5cm
$W^\pm H^\mp$ Associated Production at the Large Hadron Collider
}
\author{A. A. Barrientos Bendez\'u and B. A. Kniehl\\
{\normalsize
Max-Planck-Institut f\"ur Physik (Werner-Heisenberg-Institut),}\\
{\normalsize F\"ohringer Ring 6, 80805 Munich, Germany}\\
}

\date{}

\maketitle

\thispagestyle{empty}

\begin{abstract}
We study the production of a charged Higgs boson in association with a $W$ 
boson at the CERN Large Hadron Collider in the context of the minimal
supersymmetric extension of the standard model.
This production mechanism is particularly promising if the charged Higgs boson
is too heavy to be generated by top-quark decay.
We compare the contributions due to $b\bar b$ annihilation at the tree level
and $gg$ fusion, which proceeds at one loop.
Apart from the total cross section, we also consider distributions in 
transverse momentum and rapidity.
We also assess the viability of $W^\pm H^\mp$ associated production at the
Fermilab Tevatron after the installation of the Main Injector and the
Recycler.

\medskip

\end{abstract}

\newpage

\section{\label{sec:one}Introduction}

Despite the successful confirmation of the standard model (SM) of elementary
particle physics by experimental precision tests during the past few years,
the structure of the Higgs sector has essentially remained unexplored, and
there is still plenty of room for extensions.
A phenomenologically interesting extension of the SM Higgs sector that keeps
the electroweak $\rho$ parameter \cite{vel} at unity in the Born
approximation, is obtained by adding a second complex isospin-doublet scalar
field with opposite hypercharge.
This leads to the two-Higgs-doublet model (2HDM).
After the three massless Goldstone bosons which emerge via the electroweak
symmetry breaking are eaten up to become the longitudinal degrees of freedom
of the $W^\pm$ and $Z$ bosons, there remain five physical Higgs scalars: the
neutral CP-even $h^0$ and $H^0$ bosons, the neutral CP-odd $A^0$ boson, and
the charged $H^\pm$-boson pair.
In order to avoid flavour-changing neutral currents, one usually assumes that
all up-type fermions couple to one of the Higgs doublets while all down-type
fermions couple to the other one (2HDM of type~II).
The Higgs sector of the minimal supersymmetric extension of the SM (MSSM)
consists of such a 2HDM of type~II.
At the tree level, the MSSM Higgs sector has two free parameters, which are
usually taken to be the mass $m_{A^0}$ of the $A^0$ boson and the ratio
$\tan\beta=v_2/v_1$ of the vacuum expectation values of the two Higgs 
doublets.

The search for Higgs bosons and the study of their properties are among the 
prime objectives of the Large Hadron Collider (LHC), a proton-proton
colliding-beam facility with centre-of-mass (c.m.) energy $\sqrt S=14$~GeV 
presently under construction at CERN \cite{kun}.
At the LHC, the integrated luminosity is expected to reach $L=100$~fb$^{-1}$
per year and experiment.
In this connection, most of the attention has been focused on the neutral
Higgs bosons, and even corrections from quantum chromodynamics (QCD) to their
production cross sections and decay widths have been computed \cite{spi}.
Here, we wish to discuss the prospects of detecting $H^\pm$ bosons at the LHC.
For $H^\pm$-boson masses $m_H<m_t-m_b$, the dominant production mechanisms are
$gg,q\bar q\to t\bar t$ followed by $t\to bH^+$ and/or the charge-conjugate
decay \cite{kun}.
The dominant decay modes of $H^\pm$ boson in this mass range are
$H^+\to \bar\tau\nu_\tau$ and $H^-\to\tau\bar\nu_\tau$ unless
$\tan\beta<\sqrt{m_c/m_\tau}\approx1$ \cite{kun}.
In contrast to the SM top-quark events, this signature violates lepton
universality, a criterion which is routinely applied in ongoing $H^\pm$-boson
searches at the Fermilab $p\bar p$ collider Tevatron \cite{abe}.
For larger values of $m_H$, the most copious sources of $H^\pm$ bosons are 
provided by $gb\to tH^-$ \cite{gun,bar}, $gg\to t\bar bH^-$ \cite{dia},
$qb\to q^\prime bH^+$ \cite{mor}, and the charge-conjugate subprocesses.
The preferred decay channels are then $H^+\to t\bar b$ and $H^-\to\bar tb$,
independently of $\tan\beta$ \cite{kun}.
Unfortunately, these signal processes are bound to be obscured by large QCD
backgrounds due to $gb\to t\bar tb$, $g\bar b\to t\bar t\bar b$, and
$gg\to t\bar tb\bar b$, or by misidentification backgrounds due to
$gg,q\bar q\to gt\bar t$ and $gq\to t\bar tq$ \cite{bar}.
$H^+H^-$ pair production, which proceeds at the tree level via the
Drell-Yan process $q\bar q\to H^+H^-$, where a photon and a $Z$-boson are
exchanged in the $s$ channel \cite{eic},\footnote{%
In the case $q=b$, there are additional Feynman diagrams involving the top
quark in the $t$ channel and the $h^0$, $H^0$, and $A^0$ bosons in the $s$
channel.} and at one loop via $gg$ fusion $gg\to H^+H^-$ \cite{wil}, is also
severely plagued by such QCD backgrounds.

An attractive way out is to produce the $H^\pm$ bosons in association with
$W^\mp$ bosons, so that the leptonic decays of the latter may serve as a 
spectacular trigger for the $H^\pm$-boson search.
The dominant subprocesses of $W^\pm H^\mp$ associated production are
$b\bar b\to W^\pm H^\mp$ at the tree level and $gg\to W^\pm H^\mp$ at one 
loop.
They were numerically evaluated under LHC conditions in Ref.~\cite{dic}, for
$m_b=0$.
In this approximation, the $\bar bbh^0$, $\bar bbH^0$, and $\bar bbA^0$
couplings, which are large for $\tan\beta\gg1$, are nullified, and the
$\bar btH^-$ coupling is wrongly suppressed for $\tan\beta\gg1$.
Thus, the analysis of Ref.~\cite{dic} is only valid for $\tan\beta\approx1$.
In fact, the authors of Ref.~\cite{dic} only selected values from the interval
$0.3\le\tan\beta\le2.3$.
Furthermore, the values for $m_t$ and $\sqrt S$ and the parton density 
functions (PDF's) adopted in Ref.~\cite{dic} are now obsolete.
The purpose of this paper is to generalize the analysis of Ref.~\cite{dic} for
arbitrary values of $\tan\beta$ and to update it.
Furthermore, we shall include the leading radiative corrections to the
relations between the relevant MSSM parameters \cite{car}, which were not yet
available at the time when Ref.~\cite{dic} appeared.
In contrast to Ref.~\cite{dic}, which concentrated on the total cross section,
we shall also investigate distributions in transverse momentum $p_T$ and
rapidity $y$.
Finally, we shall also consider $W^\pm H^\mp$ associated production at the
Tevatron after the completion of the Main Injector and the Recycler (Run~II).
One expects the integrated luminosity per year and experiment then to be as
high as $L=2$~fb$^{-1}$, so that this process might provide an interesting
alternative, for moderate values of $m_H$, besides the usual $H^\pm$-production
mechanism via top-quark decay.

The literature also contains a discussion of $gg\to W^\pm H^\mp t\bar t$
\cite{kao}.
However, since the top quark turned out to be so heavy, this process is less
interesting due to the substantial phase-space suppression relative to
$gg,b\bar b\to W^\pm H^\mp$.

As for $b\bar b$ annihilation, it should be noted that the treatment of bottom
as an active flavour inside the colliding hadrons leads to an effective
description, which comprises contributions from the higher-order subprocesses
$gb\to W^\pm H^\mp b$, $g\bar b\to W^\pm H^\mp\bar b$, and
$gg\to W^\pm H^\mp b\bar b$.
If all these subprocesses are to be explicitly included along with
$b\bar b\to W^\pm H^\mp$, then it is necessary to employ a judiciously
subtracted bottom PDF in order to avoid double counting \cite{gun,kao,dad}.
The evaluation of $b\bar b\to W^\pm H^\mp$ with an unsubtracted bottom PDF is
expected to slightly overestimate the true cross section \cite{gun,kao,dad}.
For simplicity, we shall nevertheless adopt this effective approach in our
analysis, keeping in mind that a QCD-correction factor below unity is to be 
applied.

This paper is organized as follows.
In Section~\ref{sec:two}, we shall present some analytic results for the cross 
section of $W^\pm H^\mp$ associated hadroproduction via $b\bar b$
annihilation and $gg$ fusion in the 2HDM and outline our calculation of the
box amplitude.
In Section~\ref{sec:three}, we shall quantitatively analyze the size of this
cross section and estimate the number of expected signal events at the LHC and
the upgraded Tevatron.
Section~\ref{sec:four} contains our conclusions.

\section{\label{sec:two}Details of the calculation}

We start by defining the kinematics of the inclusive reaction $AB\to WH+X$,
where $A$ and $B$ are colliding hadrons, which are taken to be massless.
Let $\sqrt S$ be the energy of the initial state and $y$ and $p_T$ the 
rapidity and transverse momentum of the $W$ boson in the c.m.\ system of the
collision.
By four-momentum conservation, $m_T\cosh y\le(S+m_W^2-m_H^2)/(2\sqrt S)$, 
where $m_T=\sqrt{m_W^2+p_T^2}$ is the transverse mass of the $W$ boson.
The hadron $A$ is characterized by its PDF's $F_{a/A}(x_a,M_a)$, where $x_a$
is the fraction of the four-momentum of $A$ which is carried by the (massless)
parton $a$ ($p_a=x_a p_A$), $M_a$ is the factorization scale, and similarly
for $B$.
The Mandelstam variables $s=(p_a+p_b)^2$, $t=(p_a-p_W)^2$, and $u=(p_b-p_W)^2$
at the parton level are thus related to $S$, $y$, and $p_T$ by $s=x_ax_bS$,
$t=m_W^2-x_a\sqrt Sm_T\exp(-y)$, and $u=m_W^2-x_b\sqrt Sm_T\exp(y)$,
respectively.
Notice that $sp_T^2=tu-m_W^2m_H^2$.
In the parton model, the differential cross section of $AB\to WH+X$ is given 
by
\begin{eqnarray}
\frac{d^2\sigma}{dy\,dp_T^2}(AB\to WH+X)&=&\sum_{a,b}\int dx_adx_b\,
F_{a/A}(x_a,M_a)F_{b/B}(x_b,M_b)s\frac{d\sigma}{dt}(ab\to WH)
\nonumber\\
&&{}\times\delta(s+t+u-m_W^2-m_H^2)
\\
&=&\sum_{a,b}\int_{\bar x_a}^1dx_a\,F_{a/A}(x_a,M_a)F_{b/B}(x_b,M_b)
\frac{x_bs}{m_H^2-t}\,\frac{d\sigma}{dt}(ab\to WH),
\nonumber
\end{eqnarray}
where $\bar x_a=[\sqrt Sm_T\exp(y)-m_W^2+m_H^2]/[S-\sqrt Sm_T\exp(-y)]$ and
$x_b=[x_a\sqrt Sm_T\exp(-y)-m_W^2+m_H^2]/[x_aS-\sqrt Sm_T\exp(y)]$ in the last
expression.
The parton-level cross section is calculated from the $ab\to WH$
transition-matrix element ${\cal T}$ as
$d\sigma/dt=\overline{|{\cal T}|^2}/(16\pi s^2)$, where the average is over
the spin and colour degrees of freedom of the partons $a$ and $b$.

We now turn to the specific subprocesses $ab\to WH$.
For generality, we work in the 2HDM, adopting the Feynman rules from
Ref.~\cite{hab}.
For definiteness, however, we shall concentrate on the MSSM in the numerical
analysis in Section~\ref{sec:three}.
We neglect the Yukawa couplings of the first- and second-generation quarks.
For later use, we define here the propagator functions
\begin{eqnarray}
{\cal S}_t(s)&=&\frac{1}{\sin\beta}\left(
\frac{\cos\alpha\cos(\alpha-\beta)}{s-m_{h^0}^2+im_{h^0}\Gamma_{h^0}}
+\frac{\sin\alpha\sin(\alpha-\beta)}{s-m_{H^0}^2+im_{H^0}\Gamma_{H^0}}\right),
\nonumber\\
{\cal S}_b(s)&=&\frac{1}{\cos\beta}\left(
\frac{-\sin\alpha\cos(\alpha-\beta)}{s-m_{h^0}^2+im_{h^0}\Gamma_{h^0}}
+\frac{\cos\alpha\sin(\alpha-\beta)}{s-m_{H^0}^2+im_{H^0}\Gamma_{H^0}}\right),
\nonumber\\
{\cal P}_t(s)&=&\frac{\cot\beta}{s-m_{A^0}^2+im_{A^0}\Gamma_{A^0}},
\nonumber\\
{\cal P}_b(s)&=&\frac{\tan\beta}{s-m_{A^0}^2+im_{A^0}\Gamma_{A^0}}.
\end{eqnarray}
Here, $\alpha$ is the mixing angle that rotates the weak CP-even Higgs
eigenstates into the mass eigenstates $h^0$ and $H^0$, $m_h^0$ and
$\Gamma_{h^0}$ are the pole mass and total decay width of the $h^0$ boson,
respectively, and similarly for the $H^0$ and $A^0$ bosons.

At the tree level, $W^\pm H^\mp$ associated production proceeds via $b\bar b$
annihilation.
Here, we treat the $b$ and $\bar b$ quarks as active partons inside the
colliding hadrons $A$ and $B$.
This should be a useful picture at such high energies, $\sqrt S>m_W+m_H$.
For consistency with the underlying infinite-momentum frame, we neglect the
bottom-quark mass.
However, we must not suppress terms proportional to $m_b$ in the Yukawa 
couplings, since they generally dominate the related $m_t$-dependent terms if
$\tan\beta$ is large enough, typically for
$\tan\beta\agt\sqrt{m_t/m_b}\approx6$.
This is obvious for the $\bar btH^-$ vertex, which has the Feynman rule 
\cite{hab}
\begin{equation}
i2^{-1/4}G_F^{1/2}[m_t\cot\beta(1+\gamma_5)+m_b\tan\beta(1-\gamma_5)],
\label{eq:hbt}
\end{equation}
where $G_F$ is Fermi's constant and we have neglected the
Cabibbo-Kobayashi-Maskawa mixing, i.e., $V_{tb}=1$.
The relevant Feynman diagrams are depicted in Fig.~\ref{fig:one}.
The diagrams involving the $h^0$, $H^0$, or $A^0$ bosons are suppressed if
$\tan\beta$ is of order unity, but they are indispensable if
$\tan\beta\agt\sqrt{m_t/m_b}$.
They were neglected, along with the terms proportional to $m_b$ in 
Eq.~(\ref{eq:hbt}), in Ref.~\cite{dic}, where the restricted range
$0.3\le\tan\beta\le2.3$ was considered.
The parton-level cross section of $b\bar b\to W^-H^+$ reads
\begin{eqnarray}
\label{eq:bb}
\frac{d\sigma}{dt}(b\bar b\to W^-H^+)&=&
\frac{G_F^2}{24\pi s}\,
\left\{\frac{m_b^2}{2}\lambda\left(s,m_W^2,m_H^2\right)
\left(|{\cal S}_b(s)|^2+|{\cal P}_b(s)|^2\right)
\right.
\nonumber\\
&&{}+\frac{m_b^2\tan\beta}{t-m_t^2}\left(m_W^2m_H^2-sp_T^2-t^2\right)
\re({\cal S}_b(s)-{\cal P}_b(s))
\\
&&{}+\left.\frac{1}{\left(t-m_t^2\right)^2}
\left[m_t^4\cot^2\beta\left(2m_W^2+p_T^2\right)
+m_b^2\tan^2\beta\left(2m_W^2p_T^2+t^2\right)\right]\right\},
\nonumber
\end{eqnarray}
where $\lambda(x,y,z)=x^2+y^2+z^2-2(xy+yz+zx)$ is the K\"all\'en function.
The one of $b\bar b\to W^+H^-$ emerges through charge conjugation, by
substituting $t\leftrightarrow u$ on the right-hand side of Eq.~(\ref{eq:bb}).

An alternative $W^\pm H^\mp$ production mechanism is provided by gluon fusion,
which proceeds at one loop via the triangle-type and box diagrams depicted in
Fig.~\ref{fig:two}.
Although the parton-level cross section of gluon fusion is suppressed by two
powers of $\alpha_s$ relative to the one of $b\bar b$ annihilation, it is
expected to yield a comparable contribution at multi-TeV hadron colliders, due
to the overwhelming gluon luminosity.
On the other hand, the bottom PDF may be considered as being generated from
$g\to b\bar b$ splitting via the Altarelli-Parisi evolution and is thus of
${\cal O}(\alpha_s)$ relative to the gluon PDF.
Therefore, both mechanisms are formally of the same order at the hadron level.
As we shall see in Section~\ref{sec:three}, these two mechanisms indeed
compete with each other numerically.
Since bottom does not appear as a parton in $gg$ fusion, we keep $m_b$ finite
in this case.

The transition-matrix element of $gg\to W^-H^+$ corresponding to the sum of
the triangle-type diagrams in Fig.~\ref{fig:two} is given by
\begin{eqnarray}
{\cal T}_{\triangle}&=&\frac{\sqrt2}{\pi}\alpha_s(\mu)G_Fm_W
\varepsilon_\lambda^*(p_W)(p_a+p_b)^\lambda
\varepsilon_\mu^c(p_a)\varepsilon_\nu^c(p_b)
\nonumber\\
&&{}\times
\left[\left(p_b^\mu p_a^\nu-\frac{s}{2}g^{\mu\nu}\right)\Sigma(s)
+i\varepsilon^{\mu\nu\rho\sigma}p_{a\rho}p_{b\sigma}\Pi(s)\right],
\label{eq:tri}
\end{eqnarray}
where $\alpha_s(\mu)$ is the strong coupling constant at renormalization scale 
$\mu$, $\varepsilon_\mu^c(p_a)$ is the polarization four-vector of gluon $a$
and similarly for gluon $b$ and the $W$ boson, it is summed over the colour
index $c=1,\ldots,8$, and
\begin{eqnarray}
\Sigma(s)&=&\sum_{q=t,b}{\cal S}_q(s)S\left(\frac{s+i\epsilon}{4m_q^2}\right),
\nonumber\\
\Pi(s)&=&\sum_{q=t,b}{\cal P}_q(s)P\left(\frac{s+i\epsilon}{4m_q^2}\right).
\end{eqnarray}
Here, we have introduced the auxiliary functions
\begin{eqnarray}
S(r)&=&\frac{1}{r}\left[1-\left(1-\frac{1}{r}\right)\arsinh^2\sqrt{-r}\right],
\nonumber\\
P(r)&=&-\frac{1}{r}\arsinh^2\sqrt{-r}.
\end{eqnarray}
By analytic continuation,
$\arsinh\sqrt{-r}=-i\arcsin\sqrt r=\arcosh\sqrt r-i\pi/2$, where the first,
second, and third expressions are appropriate for $r\le0$, $0<r\le1$, and
$r>1$, respectively.
Notice that $S(r),P(r)\to0$ as $r\to\infty$, so that the bottom-quark 
contribution to Eq.~(\ref{eq:tri}) is suppressed, except for
$\tan\beta\agt m_t/m_b$.
For reference, we also list the contribution to the cross section of
$gg\to W^-H^+$ that is obtained by squaring Eq.~(\ref{eq:tri}):
\begin{equation}
\frac{d\sigma_{\triangle}}{dt}=\frac{\alpha_s^2(\mu)G_F^2}{2048\pi^3}
\lambda\left(s,m_W^2,m_H^2\right)\left(|\Sigma(s)|^2+|\Pi(s)|^2\right).
\label{eq:xs}
\end{equation}

We generated and evaluated the amplitude ${\cal T}_{\Box}$ corresponding to
the sum of the box diagrams in Fig.~\ref{fig:two} with the aid of the computer
packages Feyn Arts \cite{kue}, Feyn Calc \cite{mer}, and FF \cite{old}.
The analytic expression is somewhat lengthy, and we refrain from listing it
here.
To gain confidence in these tools \cite{kue,mer,old} and our use of them, we
checked that they allow us to numerically reproduce the differential
cross section of $gg\to ZH$ \cite{kni} in the SM to very high precision.
While finite-$m_b$ effects on ${\cal T}_{\triangle}$ are only important for
$\tan\beta\agt m_t/m_b$, such effects are indispensable in the case of
${\cal T}_{\Box}$ if $\tan\beta\agt\sqrt{m_t/m_b}$, which follows from
Eq.~(\ref{eq:hbt}).
However, neglecting $m_b$ in the bottom propagator, where it cannot be
enhanced by $\tan\beta$, should still be a useful approximation.
Nevertheless, we also keep $m_b$ finite there.
Due to Bose symmetry, the cross section $d\sigma/dt$ of $gg\to W^-H^+$ is
symmetric in $t$ and $u$.
Due to charge-conjugation invariance, it coincides with the one of
$gg\to W^+H^-$.

\section{\label{sec:three}Numerical results}

We are now in a position to explore the phenomenological implications of our
results.
The SM input parameters for our numerical analysis are
$G_F=1.16639\cdot10^{-5}$~GeV$^{-2}$ \cite{pdg} and the pole masses
$m_W=80.375$~GeV, $m_Z=91.1867$~GeV, $m_t=175.6$~GeV \cite{eww}, and
$m_b=4.7$~GeV.
We adopt the lowest-order set CTEQ4L \cite{lai} of proton PDF's.
We evaluate $\alpha_s(\mu)$ from the lowest-order formula \cite{pdg} with
$n_f=5$ quark flavours and asymptotic scale parameter
$\Lambda_{\rm QCD}^{(5)}=181$~MeV \cite{lai}.
We identify the renormalization and factorization scales with the
$W^\pm H^\mp$ invariant mass, $\mu^2=M_a^2=M_b^2=s$.
For our purposes, it is useful to select the MSSM input parameters to be
$\tan\beta$ and the pole mass $m_H$ of the $H^\pm$ bosons to be produced.
We vary them in the ranges $1<\tan\beta<40\approx m_t/m_b$ and
100~GeV${}<m_H<1$~TeV, respectively.
For given values of $\tan\beta$ and $m_H$, we determine $\alpha$ and the pole
masses $m_{h^0}$, $m_{H^0}$, and $m_{A^0}$ of the neutral Higgs bosons from
the appropriate MSSM relationships \cite{hab} including their leading
radiative corrections \cite{car} as implemented in the program package HDECAY
\cite{djo}.
In the case of $gg$ fusion, these corrections only modify
${\cal T}_\triangle$, since ${\cal T}_\Box$ does not depend on $\alpha$,
$m_{h^0}$, $m_{H^0}$, and $m_{A^0}$.
Similarly, in the case of $b\bar b$ annihilation, only the $s$-channel
diagrams are affected.
We sum over the $W^+H^-$ and $W^-H^+$ final states.

We first consider $pp\to W^\pm H^\mp+X$ at the LHC with $\sqrt S=14$~TeV.
In Fig.~\ref{fig:three}(a), the fully integrated cross sections due to
$b\bar b$ annihilation and $gg$ fusion are shown as functions of $m_H$ for
$\tan\beta=1.5$, 6, and 30.
For a comparison with future experimental data, these two contributions should
be added.
We observe that $b\bar b$ annihilation always dominates.
Its contribution modestly exceeds the one due to $gg$ fusion, by a factor of 
two or less, if $\tan\beta\agt1$ and $m_H>200$~GeV, but it is more than one
order of magnitude larger if $m_H<m_t$.
The $gg$-fusion contribution is greatly suppressed if $\tan\beta\gg6$,
independently of $m_H$.
For all values of $\tan\beta$, the latter exhibits a dip located about
$m_H=m_t$, which arises from resonating top-quark propagators in
${\cal T}_{\Box}$.
In Fig.~\ref{fig:three}(b), the $\tan\beta$ dependence of
$\sigma(pp\to W^\pm H^\mp+X)$ is displayed for $m_H=100$, 300, and 1000~GeV.
In the case of $m_H=100$~GeV, the $b\bar b$ and $gg$ contributions exhibit
minima at $\tan\beta\approx6$.
As $m_H$ increases, these minima migrate to smaller and larger values of
$\tan\beta$, respectively.
It is interesting to compare $b\bar b$ annihilation and $gg$ fusion with 
regard to their kinematic behaviour.
This is done for the $p_T$ and $y$ distributions in Figs.~\ref{fig:four}(a)
and (b), respectively, assuming $\tan\beta=1.5$, 6, 30 and $m_H=300$~GeV.
In general, the $b\bar b$ and $gg$ cross sections have similar line shapes and
just differ in their overall normalizations.
In the case of $gg$ fusion, it is instructive to analyze the interplay of
${\cal T}_\triangle$ and ${\cal T}_\Box$.
Figure~\ref{fig:five} compares the $gg$-fusion results shown in
Fig.~\ref{fig:three}(b) with the respective contributions proportional to
$|{\cal T}_\triangle|^2$ [see Eq.~(\ref{eq:xs})] and $|{\cal T}_\Box|^2$.
The latter two are comparable in size and up to one order of magnitude larger
than the full result.
Obviously, there is a strong destructive interference between
${\cal T}_{\triangle}$ and ${\cal T}_{\Box}$.
For a typical MSSM scenario \cite{djo} with $\tan\beta$ and $m_H$ in the
ranges considered here, the relative shift in the $gg$ ($b\bar b$) cross
section due to the MSSM radiative corrections \cite{car} does not exceed the
order of 10\% (1\%) in magnitude.

As advertized in Section~\ref{sec:one}, one of the phenomenological advantages 
of $W^\pm H^\mp$ associated production is the circumstance that the charged 
leptons originating from the decaying $W^\pm$ bosons can be utilized as a
clean trigger.
Isolated, energetic electrons and muons will be hard to miss, and $\tau$
leptons should be identifiable with high efficiency via their one-prong decays
to electrons, muons, charged pions, or charged kaons, which have a combined
branching fraction of about 85\% \cite{pdg}.
Thus, approximately 30\% of the $W^\pm H^\mp$ signal events should be more or 
less straightforwardly detectable in this way.
If we assume the integrated luminosity per year to be at its
design value of $L=100$~fb$^{-1}$ for each of the two LHC experiments, ATLAS 
and CMS, then a cross section of 1~fb translates into about 60 detectable
$W^\pm H^\mp$ events per year.
Looking at Fig.~\ref{fig:three}, we thus conclude that, depending on 
$\tan\beta$, one should be able to collect an annual total of between 650 and
14,000 such events if $m_H=300$~GeV.

We now turn to $p\bar p$ collisions at the Tevatron with $\sqrt S=2$~TeV
(Run~II).
In Fig.~\ref{fig:six}, the total cross sections due to $b\bar b$ annihilation
and $gg$ fusion are presented as functions of $m_H$ for $\tan\beta=1.5$, 6,
and 30.
During Run~II, the Tevatron, supplemented by the Main Injector and the 
Recycler, is expected to deliver an integrated luminosity of $L=2$~fb$^{-1}$
per year to each of the two detectors, CDF and D0.
Assuming that the $H^\pm$ bosons can be identified via their decays to $\tau$ 
leptons and that the $W^\pm$ bosons can also be recognized if they decay
hadronically, by requiring that the two-jet invariant mass be close to $m_W$,
a cross section of 1~fb hence corresponds to about 20 detectable $W^\pm H^\mp$
events during five years of operation.
From Fig.~\ref{fig:six}, we read off that, depending on $\tan\beta$, the total
yield during that period should range between 5 and 50 if $m_H=100$~GeV.

Finally, we should compare our analysis with the one reported in
Ref.~\cite{dic}.
If we adopt the input information from Ref.~\cite{dic}, we are able to nicely
reproduce the results obtained therein, except that our $gg$-fusion cross
section turns out to be a factor of two larger.
A possible interpretation of this difference is that, in contrast to the case 
of $b\bar b$ annihilation, the results for $gg$ fusion shown in Figs.~4 and 5
of Ref.~\cite{dic} actually refer to one of the $W^+H^-$ and $W^-H^+$ final
states rather than to their sum as declared in the text.

\section{\label{sec:four}Conclusions}

\begin{figure}[ht]
\begin{center}
\centerline{\epsfig{figure=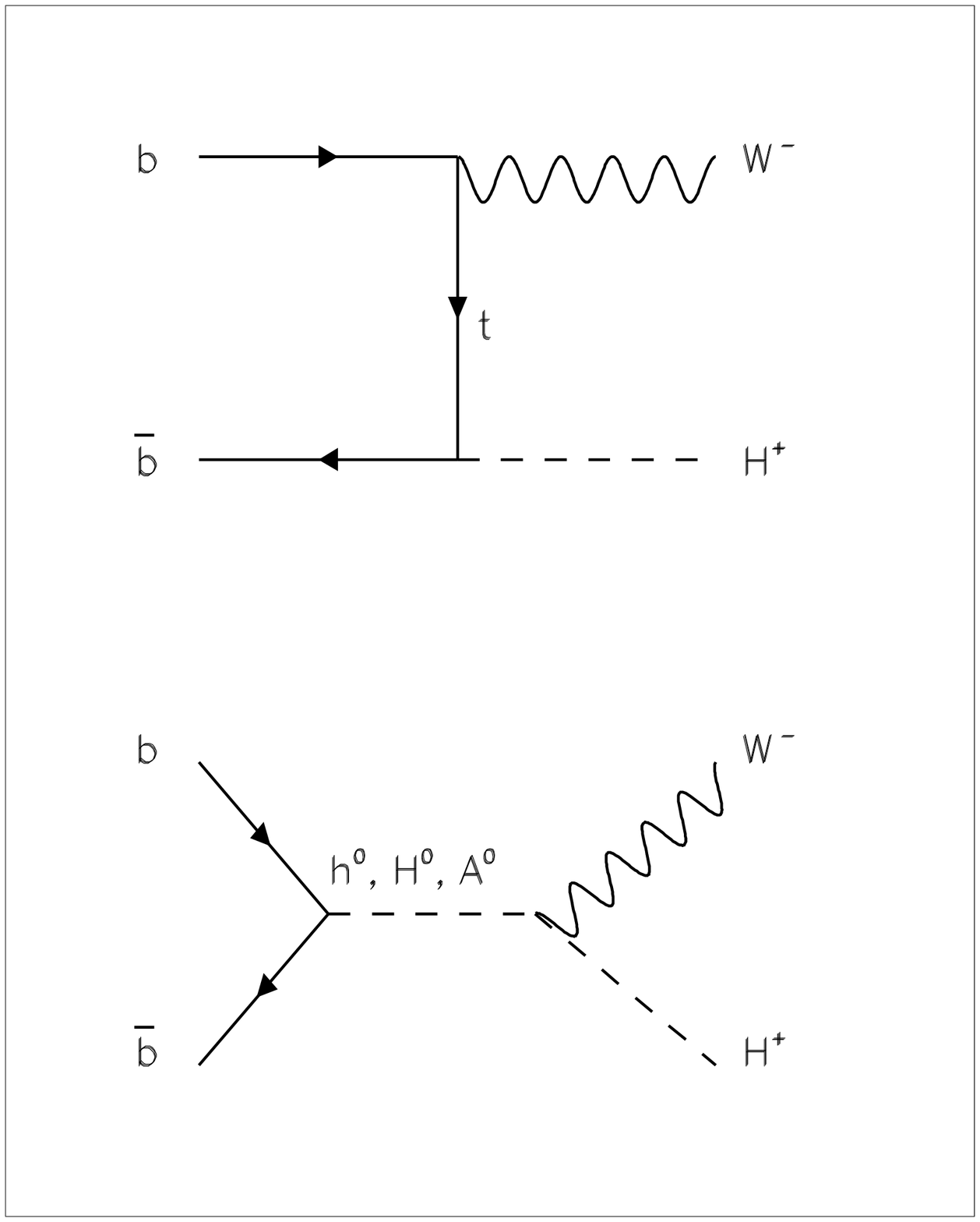,height=18cm}}
\caption{Feynman diagrams for $b\bar b\to W^-H^+$.}
\label{fig:one}
\end{center}
\end{figure}

\begin{figure}[ht]
\begin{center}
\centerline{\epsfig{figure=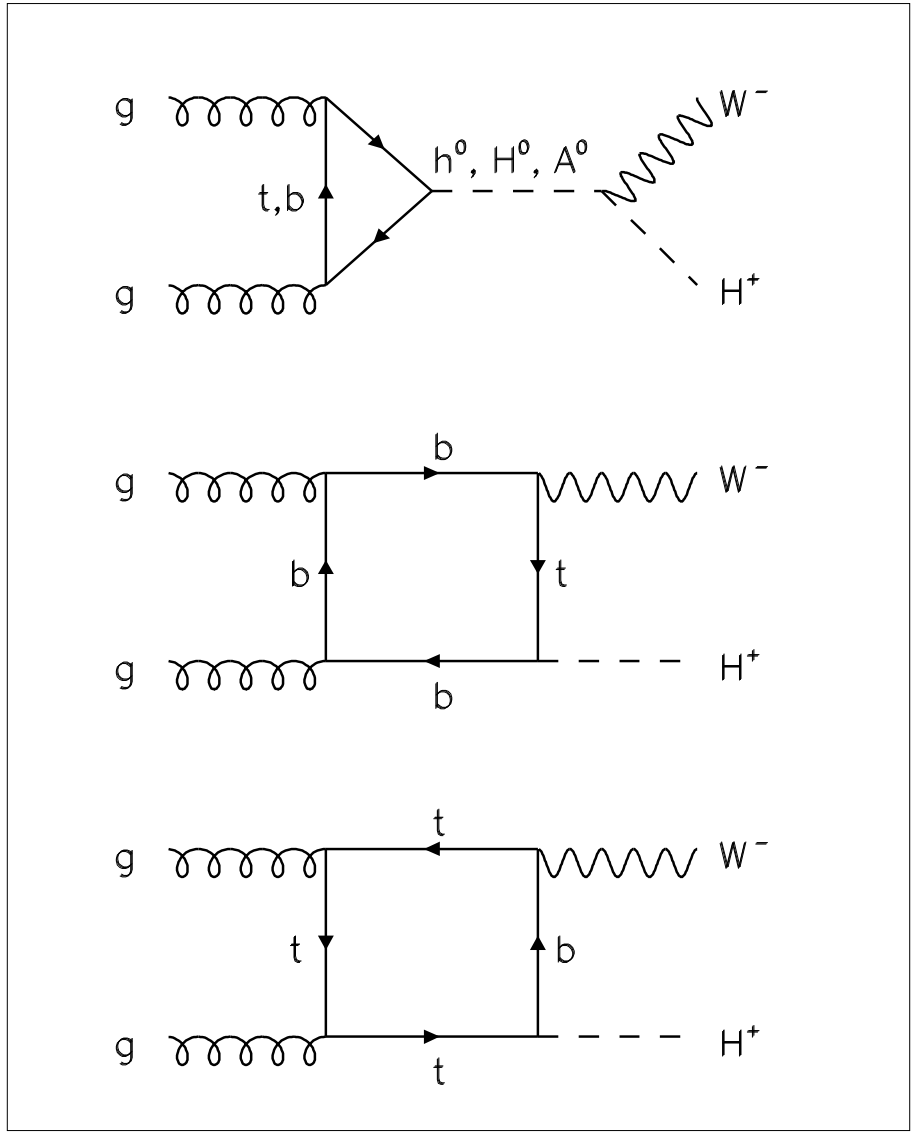,height=18cm}}
\caption{Typical Feynman diagrams for $gg\to W^-H^+$.}
\label{fig:two}
\end{center}
\end{figure}

\begin{figure}[ht]
\begin{center}
\begin{tabular}{cc}
\parbox{8cm}{\epsfig{figure=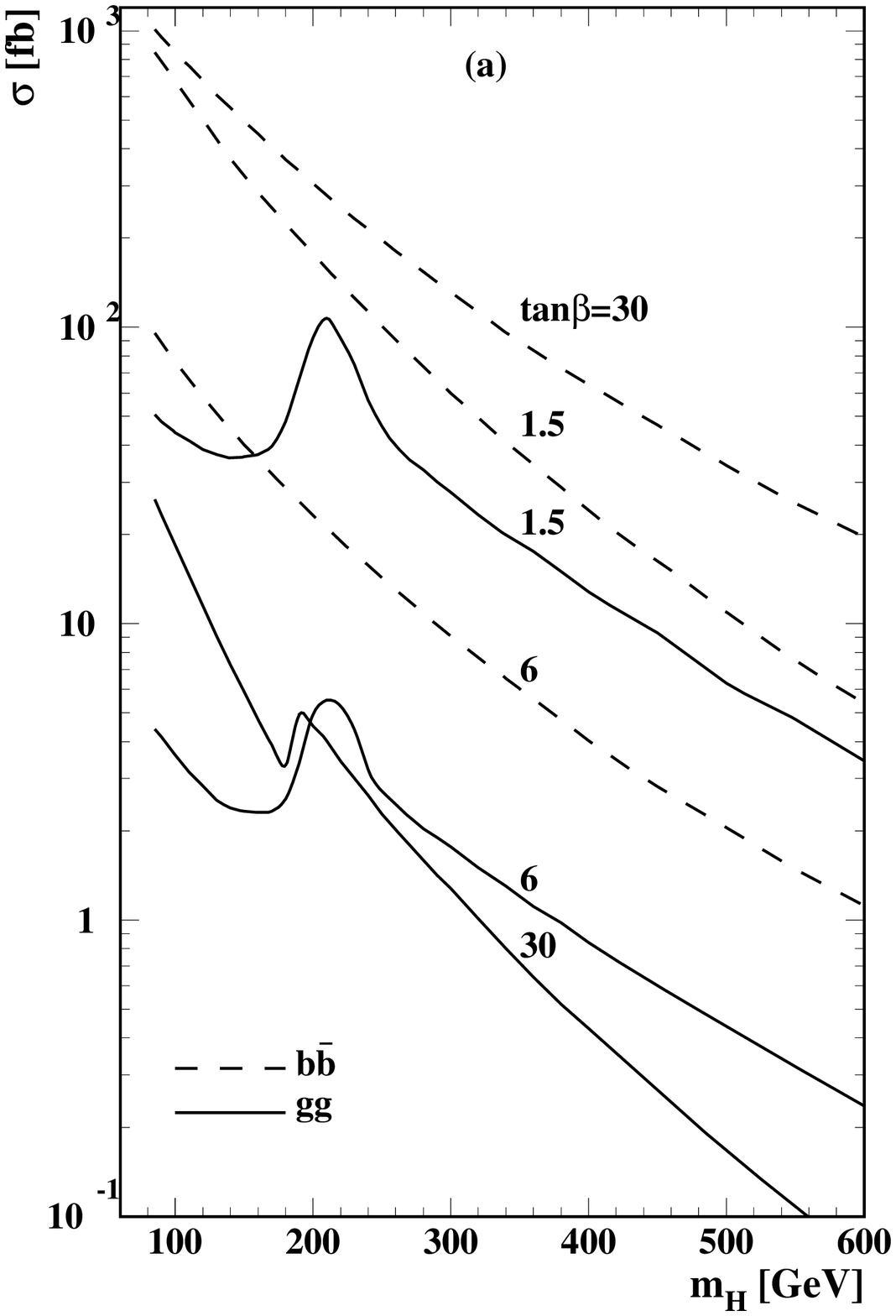,width=8cm}}
&
\parbox{8cm}{\epsfig{figure=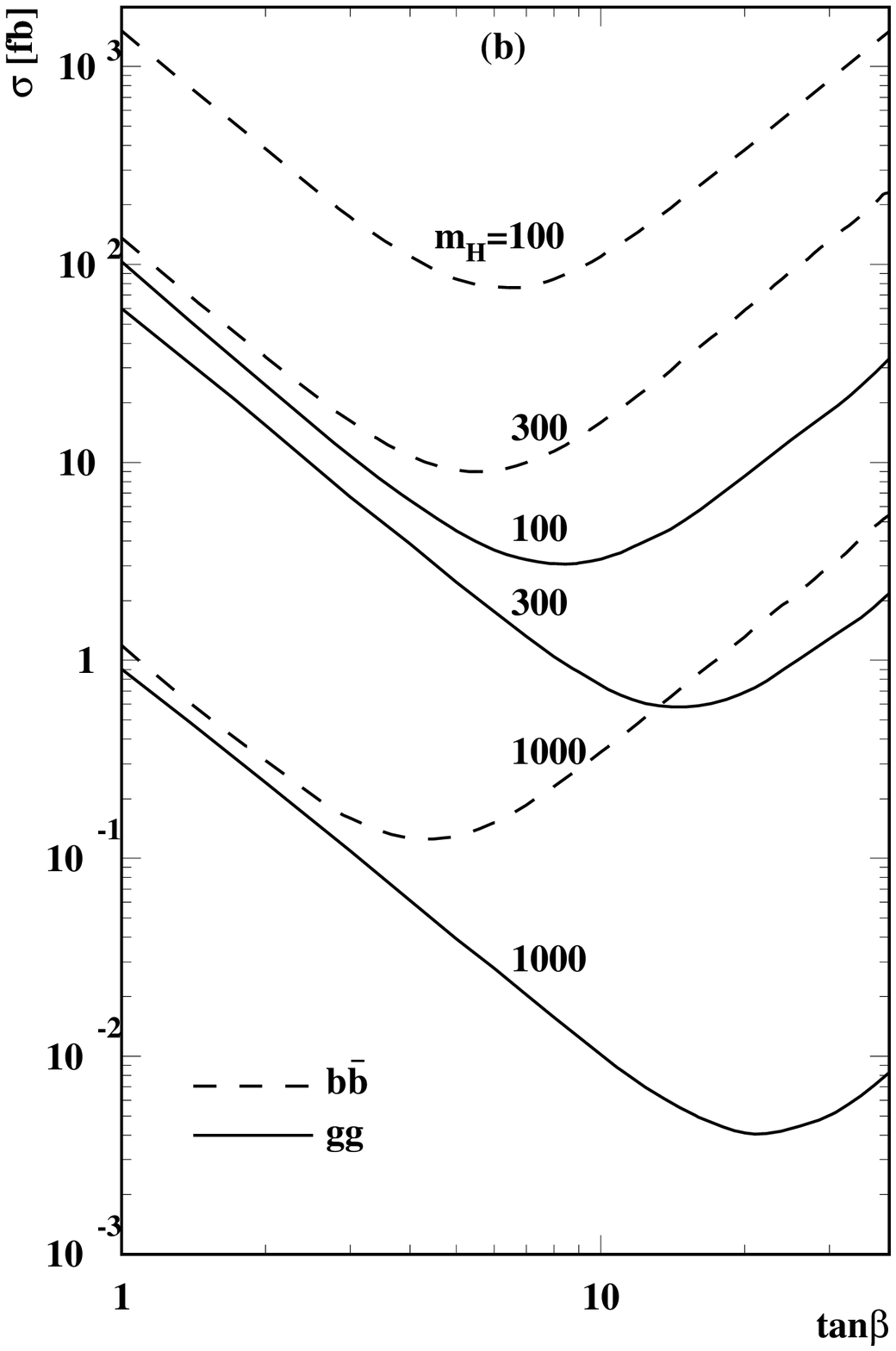,width=8cm}}
\end{tabular}
\caption{Total cross sections $\sigma$ (in fb) of $pp\to W^\pm H^\mp+X$ via
$b\bar b$ annihilation (dashed lines) and $gg$ fusion (solid lines) at the LHC
(a) as functions of $m_H$ for $\tan\beta=1.5$, 6, and 30; and (b) as functions
of $\tan\beta$ for $m_H=100$, 300, and 1000~GeV.}
\label{fig:three}
\end{center}
\end{figure}

\begin{figure}[ht]
\begin{center}
\begin{tabular}{cc}
\parbox{8cm}{\epsfig{figure=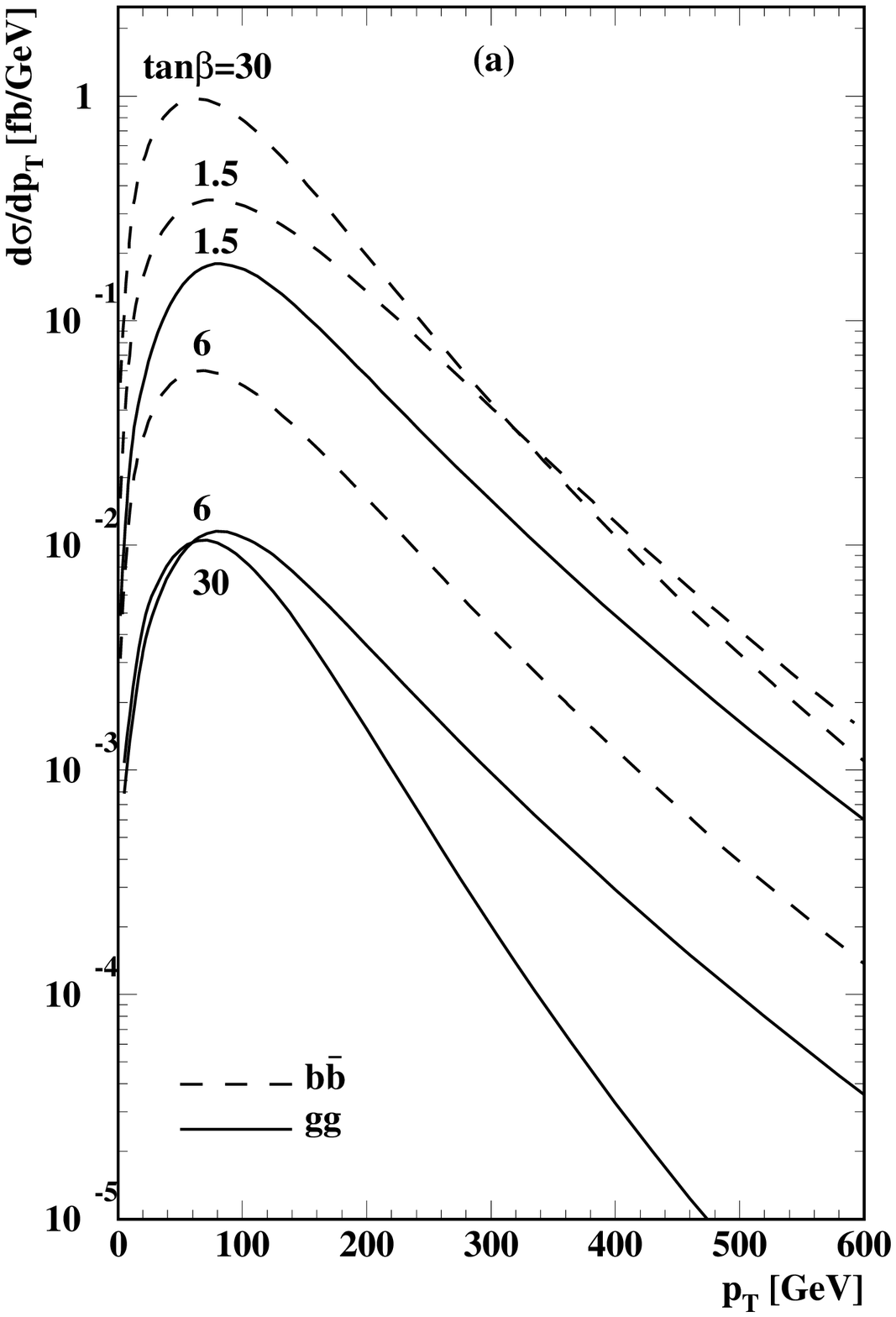,width=8cm}}
&
\parbox{8cm}{\epsfig{figure=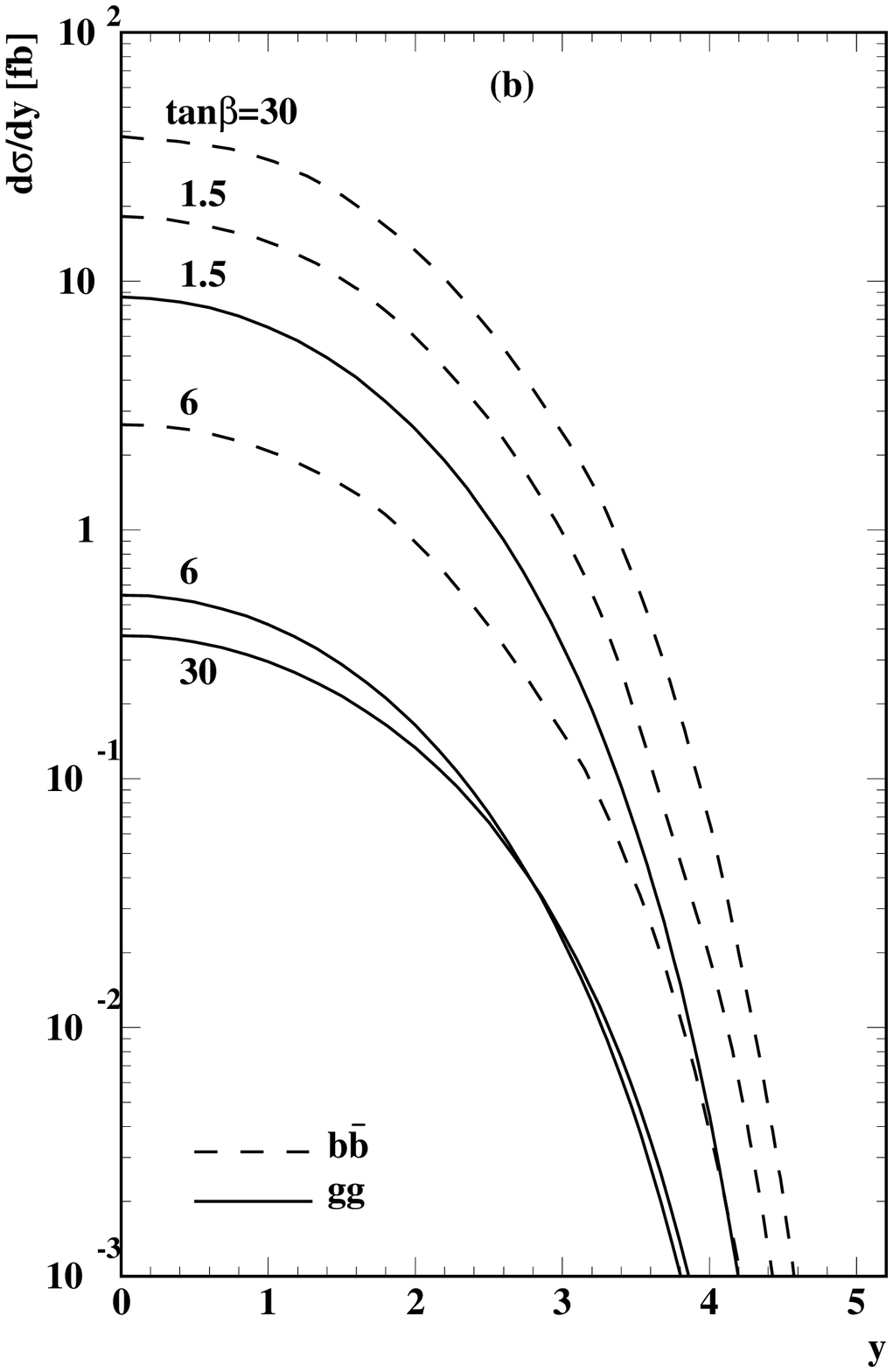,width=8cm}}
\end{tabular}
\caption{(a) $p_T$ distributions $d\sigma/dp_T$ (in fb/GeV) and (b) $y$ 
distributions $d\sigma/dy$ (in fb) of $pp\to W^\pm H^\mp+X$ via $b\bar b$
annihilation (dashed lines) and $gg$ fusion (solid lines) at the LHC for
$\tan\beta=1.5$, 6, 30 and $m_H=300$~GeV.}
\label{fig:four}
\end{center}
\end{figure}

\begin{figure}[ht]
\begin{center}
\centerline{\epsfig{figure=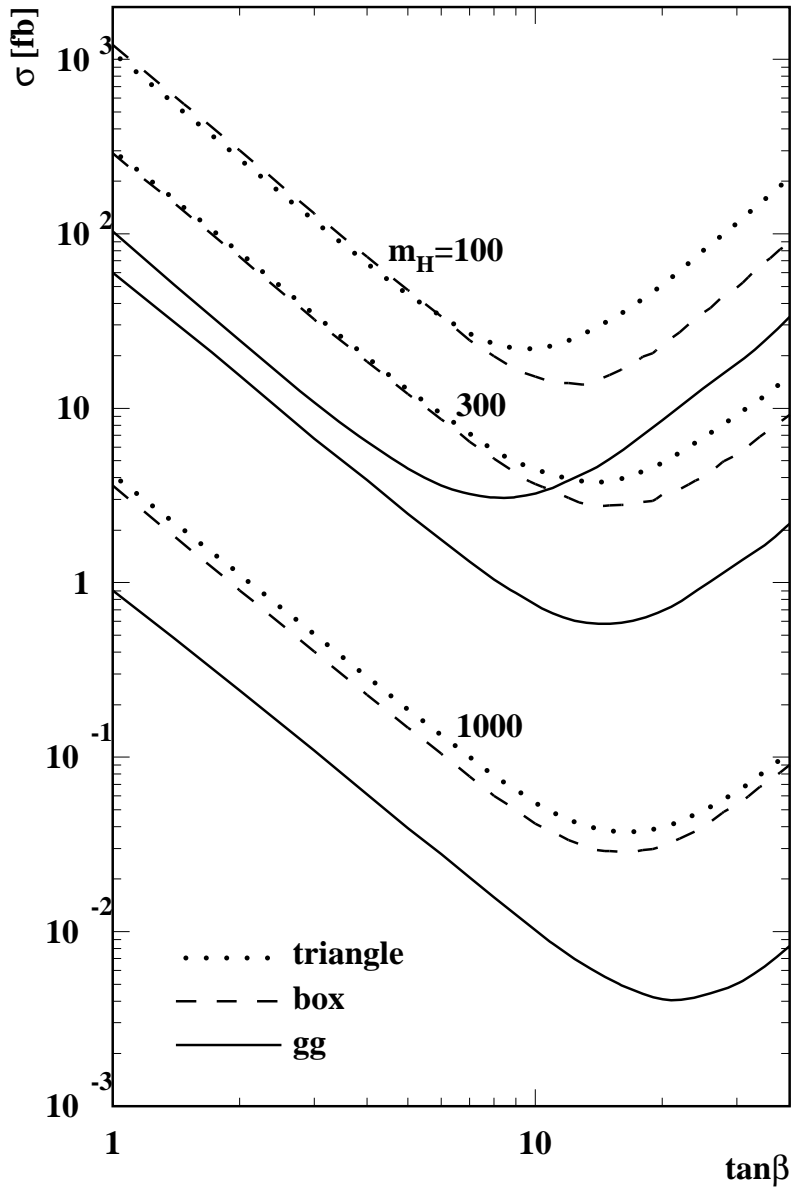,height=18cm}}
\caption{Total cross section $\sigma$ (in fb) of $pp\to W^\pm H^\mp+X$ via
$gg$ fusion (solid lines) at the LHC as a function of $m_H$ for
$\tan\beta=1.5$, 6, and 30.
The contributions due to the triangle-type (dotted lines) and box diagrams
(dashed lines) are also shown.}
\label{fig:five}
\end{center}
\end{figure}

\begin{figure}[ht]
\begin{center}
\centerline{\epsfig{figure=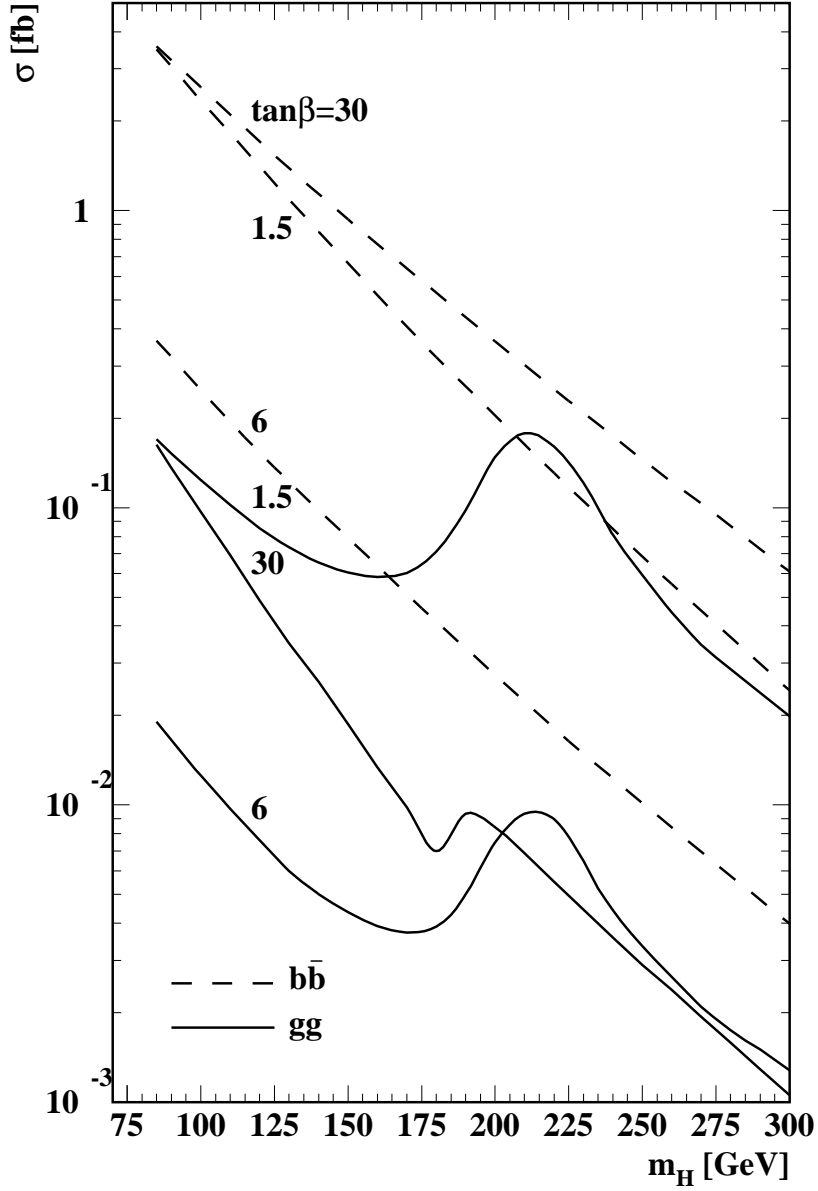,height=18cm}}
\caption{Total cross sections $\sigma$ (in fb) of $p\bar p\to W^\pm H^\mp+X$
via $b\bar b$ annihilation (dashed lines) and $gg$ fusion (solid lines) at the 
Tevatron (Run~II) as functions of $m_H$ for $\tan\beta=1.5$, 6, and 30.}
\label{fig:six}
\end{center}
\end{figure}

We studied $W^\pm H^\mp$ associated hadroproduction within the MSSM, allowing 
for $\tan\beta$ to be arbitrary.
We included the contributions from $b\bar b$ annihilation and $gg$ fusion to
lowest order.
For $\tan\beta\agt6$, the $m_b$-dependent terms in the relevant Yukawa
couplings give rise to significant effects in both channels and must not be
neglected.
In particular, the $s$-channel diagrams of Fig.~\ref{fig:one} would otherwise
be missed.
We also incorporated the leading corrections to the relations between the
relevant MSSM parameters \cite{car}.

Using up-to-date information on the input parameters and proton PDF's, we
presented theoretical predictions for the $W^\pm H^\mp$ production cross
section at LHC and Tevatron energies.
Apart from the fully integrated cross section, we also analyzed distributions 
in $p_T$ and $y$.
A favourable scenario for $W^\pm H^\mp$ associated hadroproduction would be 
characterized by the conditions that $m_H>m_t-m_b$ and that $\tan\beta$ is 
either close to unity or of order $m_t/m_b$.
Then, the $H^\pm$ bosons could not spring from on-shell top quarks, which are
so copiously produced at hadron colliders, and their decays to $\tau$ leptons,
which are relatively easy to identify, would have a small branching fraction.
On the other hand, $W^\pm H^\mp$ production would have a sizeable cross
section, and the leptonic $W^\pm$ decays would provide a spectacular trigger.
We found that the $W^\pm H^\mp$ signal should be clearly visible at the LHC
unless $m_H$ is very large.
The search for this signal could also usefully supplement the standard 
techniques of looking for $H^\mp$ bosons \cite{abe} during Run~II at the
Tevatron.

\vspace{1cm}
\noindent
{\bf Acknowledgements}
\smallskip

\noindent
We thank Peter Zerwas for suggesting this project, to Sally Dawson, Karl
Jakobs, and Gordon Kane for instructive discussions, to Thomas Hahn and
Georg Weiglein for useful advice regarding the implementation and operation of
Feyn Arts \cite{kue} and Feyn Calc \cite{mer}, and to Michael Spira for a
helpful remark concerning Ref.~\cite{djo}.
The work of A.A.B.B. was supported by the Friedrich-Ebert-Stiftung through
Grant No.~219747.

\newpage

\newpage

\newpage

\newpage

\newpage

\newpage

\end{document}